\documentclass[%
 aip,
 amsmath,amssymb,
 reprint,%
]{revtex4-1}

\usepackage{graphicx}
\usepackage{dcolumn}
\usepackage{bm}

\usepackage[latin1,utf8]{inputenc}
\usepackage[T1]{fontenc}
\usepackage{textcomp}
\usepackage{mathptmx}

\begin{document}

\preprint{AIP/123-QED}

\title{\LARGE \bf
Ion Mobility Independent Large Signal Switching of Perovskite Devices 
}

\author{Saketh Tirupati, Abhimanyu Singareddy, Dhyana Sivadas, and Pradeep R. Nair$^*$\\
Department of Electrical Engineering, Indian Institute of Technology Bombay, Mumbai, India\\
E-mail: prnair@ee.iitb.ac.in
}

\begin{abstract}
 The presence of mobile ions in perovskites is well known to influence the device electrostatics leading to a wide variety of anomalous characteristics related to hysteresis, efficiency degradation, low frequency capacitance, large signal switching, etc. Accordingly, the ion mobility is understood to a have a critical influence on the associated time constants/delays. Quite contrary to this broadly accepted thought, here we show that the time delays associated with large signal switching show a universal behavior dictated by electronic dipoles, rather than ionic dipoles. Due to the resultant sudden and dramatic collapse of contact layer depletion region, switching delays are independent of ion mobilities! Further, our detailed numerical simulations, well supported by experimental results, indicate that terminal currents show near steady state behavior well ahead of the relaxation of ionic distributions to their steady state conditions. These results have interesting implications towards the understanding and optimization of perovskite based electronic devices, including solar cells and LEDs.\\
\newline
\textit{keywords: Ion migration, hysteresis, drift-diffusion}
\end{abstract}

\maketitle

\section{Introduction}
\begin{figure}[ht]
  \centering
    \includegraphics[width=0.45\textwidth]{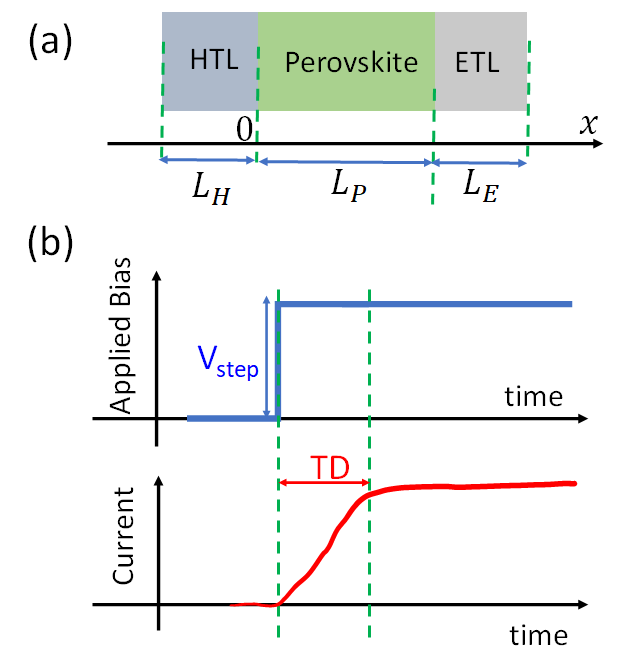}
    \caption{\textit{Large signal switching of perovskite PIN devices. (a) Schematic of the device. Here $L_H$, $L_P$, and $L_E$ denote the thickness of HTL, perovskite, and ETL, respectively. (b) Schematic of applied bias, and resultant current through the device depicting the time delay (TD).}}
\label{Fig1}
\end{figure}

Ion migration and related aspects are among the most explored topics related to the long term stability of perovskite solar cells. Multiple ion species with a wide range of mobilities are understood to be present in the perovskite active material\cite{park_apl_2019}. For example, the reported ionic densities vary from $10^{17}cm^{-3}$ to $10^{20}cm^{-3}$. Similarly, the reported mobility ($\mu_{I}$) of the dominant ionic species range from $10^{-8}cm^2/Vs$ to $10^{-14}cm^2/Vs$. Such a broad range of ionic density and mobility along with surface/interface recombination contributes to phenomena ranging from efficiency degradation to hysteresis\cite{reenen_jpcl,koster_jpcc,weber_ees}.
\vspace{0.5em}\\
A key impact of the presence of low mobility ions at large densities  is that the device often undergoes a lengthy transient phase - in contrast to the very short settling times routinely observed in devices based on inorganic semiconductors\cite{park_adm}. One such scenario is depicted in Fig. \ref{Fig1}, where the device subjected to a large signal switching displays a  time delay ($TD$) in current response.
A first order analysis indicate that $TD$ under diffusion dominant transport is given as $TD_{diff} \sim W^{2}/(\mu_IV_T)$, where $W$ is the relevant spatial extent over which ionic transport is crucial, and $V_T=kT/q$, the thermal voltage. On the other hand, under the influence of electric field, we have $TD_{drift} \sim W^2/(\mu_{I}\Delta V)$, where $\Delta V$ is the potential drop that drives drift of ions across $W$. Accordingly, regardless of the underlying transport mechanism, we obtain a simple scaling relation for $TD$ as
\begin{equation}
    TD \propto \mu_{I}^{-1}.
\end{equation}
A broad range of $TDs$ are reported in literature \cite{C6TC04964H} \cite{doi:10.1021/acsami.9b04991} which do not conform to the above analysis. Unravelling this puzzle through detailed numerical simulations, here we show that the $TD$ is independent of $\mu_I$, for large signal switching beyond $V_{bi}$ (see Fig. \ref{Fig2}), in contrast to the scaling relation, eq (1). This surprising trend has its origins on electronic charge build up (i.e., even beyond that of steady state values) and an associated dramatic collapse of contact layer/perovskite depletion region - a phenomena not yet explored/understood in the perovskite community.\\

Rest of manuscript is devoted to a detailed exploration of device physics involved in large signal switching of  perovskite devices. Below we first describe the model system and constituent equations, followed by the numerical simulation results.\\

\section{Model System}
 A schematic of the device under study is provided in Fig. \ref{Fig1}a.  It consists of a classical PIN architecture where the photo-active layer (i.e., perovskite) is sandwiched between the electron transport layer (ETL) and hole transport layer (HTL). The ETL and HTL are heavily doped while the perovskite is undoped. Large signal switching of the device is depicted in Fig. \ref{Fig1}b. Here, the device is switched, under dark conditions, from $V=0$ to a large forward bias $V_{step}$. The time delay in diode current (defined as the time taken for the current to settle to $80\%$ of its steady value) is the parameter of main interest to this work.\vspace{0.5em}\\
 To explore the effects of ion migration, the semiconductor device equations are numerically solved. Specifically, we aim at self-consistent solution of Poisson's equation for electrostatics and continuity equations for electrons, holes, and mobile ions. For ease of reference, the relevant equations are provided below for the perovskite region. Similar equations with appropriate parameters describe the carrier transport in ETL and HTL.
 
 \begin{equation}
     \epsilon\frac{\partial^2 \Psi}{\partial^2 x}=-q(\, P - N + N_{I,F}-N_{I,M}  )\,
 \end{equation}
  \begin{equation}
    \frac{\partial N}{\partial t}=\frac{1}{q}\frac{\partial J_N}{\partial x}-r_{N}+g_{N}
 \end{equation}
   \begin{equation}
    \frac{\partial P}{\partial t}=-\frac{1}{q}\frac{\partial J_P}{\partial x}-r_{P}+g_{P}
 \end{equation}
    \begin{equation}
    \frac{\partial N_{I,M}}{\partial t}=\frac{1}{q}\frac{\partial J_{I,M}}{\partial x}
 \end{equation}
 Here $\Psi$, $N$, and $P$ denote the electrostatic potential, electron density and hole density, respectively. Also the parameters $x$ and $t$ denote the spatial coordinate and time, respectively (rest of the parameters have their usual notation and are defined in Suppl. Info). The current density $J$ for each species is described through the drift-diffusion formalism. In accordance with earlier reports, we consider two ionic species confined to the perovskite active layer - (i) a positively charged immobile species ($N_{I,F}$) uniformly distributed over the entire perovskite region, and (ii) negatively charged mobile species ($N_{I,M}$). In addition, we assume that the net ionic charge in the perovskite active material is zero - i.e.,  $\int_{0}^{L_P}\! N_{I,M}(t) \, \mathrm{d}x=N_{I,F}L_P$ (where $L_P$ is the thickness of perovskite layer, see Fig. \ref{Fig1}a). We observe that there is significant debate in literature on the polarity of mobile ions. In this context, as our main focus is about the role of electrostatics on transient delays, the choice of polarity of mobile ion species is well justified. \vspace{0.5em}\\ 
 The above set equations were self consistently solved in 1 dimension through finite difference method with  Scharfetter-Gummel discretization, backward Euler method for time integration, and Newton's method for iterative solution of coupled non-linear equations. Definition of rest of the variables, parameters used, and the detailed simulation methodology is provided in Suppl. Materials. 
 
 \section{Results}
\begin{figure}[ht]
  \centering
    \includegraphics[width=0.45\textwidth]{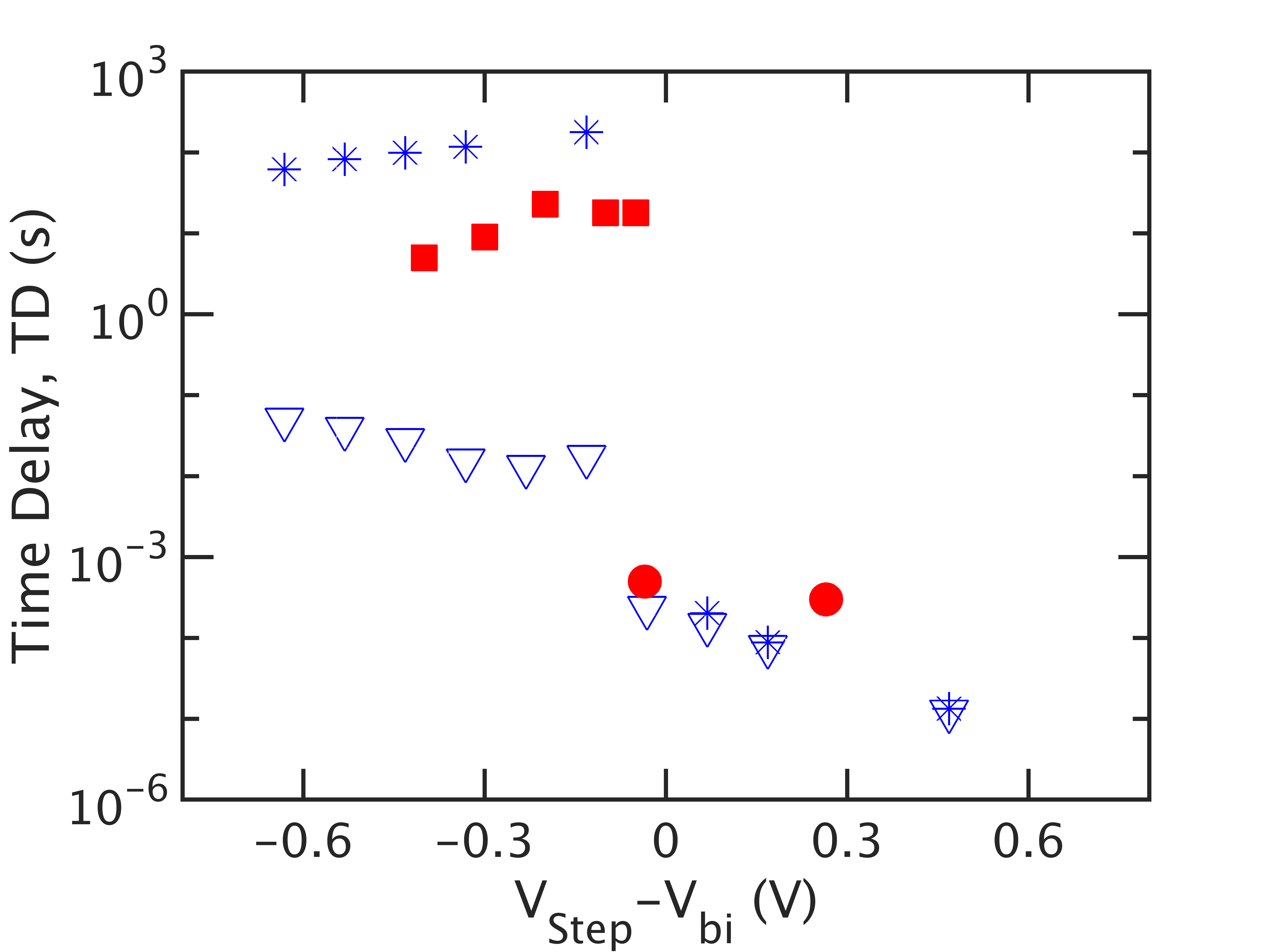}
    \caption{\textit{ Variation of TD as a function of applied bias. Open symbols denote simulation results (this work, open triangles: $\mu_I=10^{-8}cm^2/Vs$, and asterisks:$\mu_I=10^{-12}cm^2/Vs$) while solid symbols denote results from literature. The results indicate that for $V_{step}>V_{bi}$, $TD$ is independent of $\mu_I$.}}
\label{Fig2}
\end{figure}

 Figure \ref{Fig2} summarizes the main results of this manuscript. Here the TD is plotted against  $V_{step}-V_{bi}$. For simulations, $V_{bi}=1.03V$ as per the set of parameters used. Closed symbols represent data from literature - square symbols from ref. \cite{C6TC04964H}  (with $V_{bi}=1.06V$, as mentioned in the same reference) while dots are from ref. \cite{doi:10.1021/acsami.9b04991} (with $V_{bi}=1.23V$, as estimated using the parameters provided in same reference). The $TD$ associated with large signal switching reported in ref \cite{C6TC04964H} is of the order of 10s with $\mu_I\approx 10^{-14} cm^2/Vs$. On the other hand, ref. \cite{doi:10.1021/acsami.9b04991} reported delays of the order of 1ms with $\mu_I\approx10^{-8}cm^2/Vs$. Eq. (1) anticipates that the ratio of $TDs$ of these two independent results should be around $10^{6}$, while the ratio of the TDs experimentally observed is around $10^4$ - a mismatch of two orders of magnitude. We note that as both the above reports had multiple data sets with similar $TD$, the mismatch from predictions of eq. (1) is more likely to be due to certain unexplained fundamental aspects rather than any random effects.\vspace{-0.5em}\\ 

\begin{figure}[ht]
  \centering
    \includegraphics[width=0.45\textwidth]{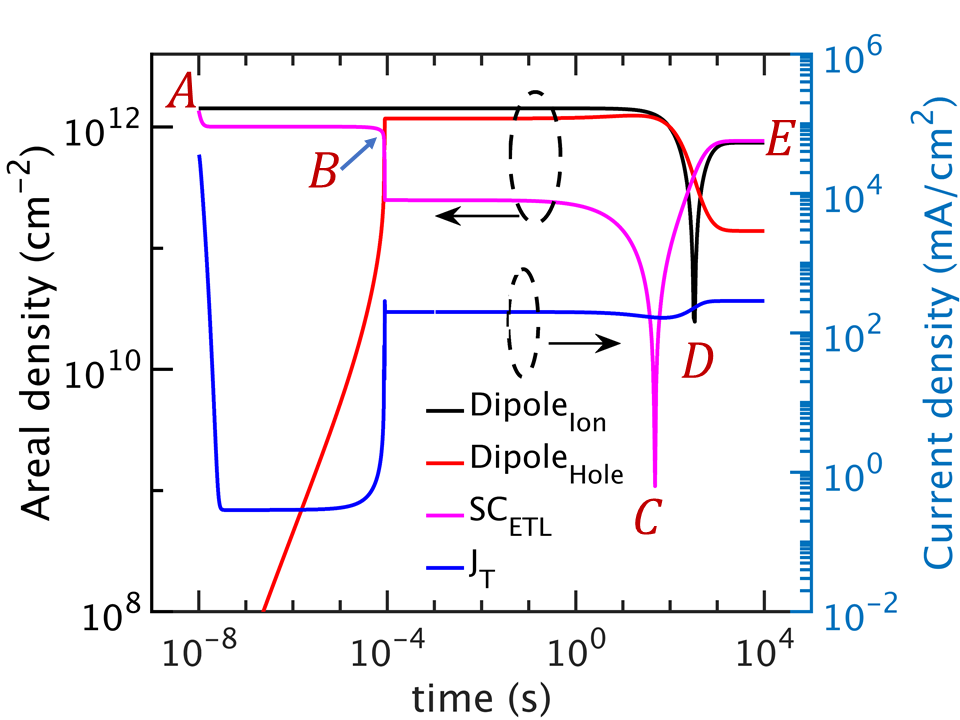}
    \caption{\textit{ Current transient (plotted to the right Y axis) in response to large signal switching ($V_{step}=1.2V$). Time evolution of various charge components (plotted to the left Y axis) indicate that accumulated carriers ($Dipole_{Hole}$) trigger a sudden collapse of depletion region in the contact layers (i.e., event B).  }}
\label{Fig4}
\end{figure}

 Simulation results shared in Fig. \ref{Fig2} correspond to an average mobile ion density of $10^{18}cm^{-3}$ and for two different ion mobilities - (a) $\mu_I=10^{-8}cm^2/Vs$ (shown by triangles), and (b) $\mu_I=10^{-12}cm^2/Vs$ (shown by asterisk symbols). Clearly for $V_{step}<V_{bi}$, the time delays almost scale as per eq (1) indicating that the ionic motion play a major role. Surprisingly, the simulation results for $V_{step}>V_{bi}$ indicate that the time delays are almost independent of $\mu_I$. Interestingly, the data from ref. \cite{doi:10.1021/acsami.9b04991} (dots) are in this high bias regime and hence have no strong dependence on the ion mobility. This unravels the puzzle as to why the experimentally observed time delays do not scale with mobility - that $TDs$ are strongly influenced by the bias conditions and eq. (1) is no longer applicable for $V_{step}> V_{bi}$. We now systematically explore the physical reason behind this unexpected result.\vspace{0.5em}\\
\textbf{Steady State:} The band diagram and the profiles of $N_{I,M}$, $N$, and $P$ under equilibrium and under steady state conditions with $V_{step}=1.2V$ are provided in Suppl. Mat (see Fig. \ref{Fig3}). Negative mobile ions accumulate near the ETL side at equilibrium. Consequently, the space charge near the HTL side is dominated by the positive immobile ions. Since the applied bias $V_{step} = 1.2V > V_{bi}$, negative mobile ions accumulate near the HTL side under steady state conditions. Note that under such conditions, electrons and holes accumulate near HTL and ETL interfaces, respectively (see Fig. \ref{Fig3}b).\vspace{0.5em}\\
It is evident that a large signal switching  from $0V$ to $V_{step}>V_{bi}$ entails at least 3 changes - (i) flipping of ionic dipole, (ii) build up electrons and holes in the active region, and (iii) flipping of the bands at the interfaces. We note that these changes may or may not happen concurrently. The order of occurrence of the same and any additional important events could determine the eventual current transients. Further, it is instructive to note that the perovskite active region has two dipoles - one due to the ionic charge and another due to the accumulated electrons and holes in the perovskite region. Interestingly, the correlated dynamics of these dipoles and their relative magnitude are instrumental in mobility independent fast switching, as explained below. \vspace{0.5em} \\ 
\textbf{Transient events:} Time variation of current in response to the large signal switching is plotted in Fig. \ref{Fig4}. It has several regimes of interest. Broadly, the current transient consist of an initial decay followed by a sharp increase in current which gradually reach the steady state value. We note that the broad trends of the current transient are consistent with the literature \cite{doi:10.1021/acsami.9b04991}. To gain further information on the phenomena that dictates the features of transient curve, the areal density of net ionic charge ($Dipole_{Ion}=\int_{L_P/2}^{L_P}\! (N_{I,M}-N_{I,F}) \, \mathrm{d}x$) and holes ($Dipole_{Hole}=\int_{L_P/2}^{L_P}\! P \, \mathrm{d}x$) in the perovskite region near the ETL interface is plotted in the same figure. In addition, the space charge in ETL ($SC_{ETL}=\int_{L_P}^{L_P+L_E}\! (N-N_{D}) \, \mathrm{d}x$) is also plotted which gives information on the time evolution of the band bending or depletion region in ETL. 
The parameters $L_P$ and $L_E$ are defined in Fig. \ref{Fig1}a and $N_D$ is the ETL doping density. Note that electrons accumulate next to HTL interface and hence is not plotted in Fig. \ref{Fig4}. In fact, the areal density of electrons accumulated near HTL region is very similar to $Dipole_{Hole}$ plotted in Fig. \ref{Fig4}. \vspace{0.5em}\\
Important features of the transient current can be understood through analysis of various discrete time events depicted in Fig. \ref{Fig4}. Note that $J_T$ corresponds to the total transient current through the device. The $J_T$  plotted in Fig. \ref{Fig4} was evaluated in the quasi-neutral region of the ETL (or HTL as well). Since the electric fields are negligible, the current in ETL is entirely due to the motion of electrons (with negligible displacement current components). 
\vspace{0.5em}\\
The initial conditions for the transient simulations are equilibrium distribution of $N$, $P$ and $\Psi$ for $V=0$. The E-B diagram under such conditions is provided in Fig. \ref{Fig3}a. Due to the electrostatic screening by ions, the electric field is negligible over most of the perovskite region. Consequently, at moment A (see Fig. \ref{Fig4}), i.e., at the start of transient simulations, the ionic charge is exactly matched by the space charge in ETL. Sudden application of $V_{step}=1.2V$ results in a reduction of band bending and hence a reduction in the depletion region of ETL (see the rapid drop in $SC_{ETL}$ near event A, Fig. \ref{Fig4} ). However, note that the depletion region do not entirely disappear in the ETL side. \vspace{0.5em}\\
Once the depletion layer in ETL responds to $V_{step}$ (i.e., after event A), the current becomes injection limited and remains nearly a constant till the event B. Over this duration, electrons and holes accumulate near the HTL and ETL regions, respectively. For convenience, only the hole accumulation near the HTL is plotted in Fig. \ref{Fig4}. Negligible carrier recombination happens during this phase as the carrier densities are not significant (further they are localized and spatially spread apart). As the $Dipole_{Hole}$ become comparable to the $Dipole_{Ion}$ at ETL side, a sudden collapse of the ETL depletion region occurs which results in a sharp increase in current (Event B, see Fig. \ref{Fig4}). Note that the ionic dipole (i.e., $Dipole_{Ion}$ curve) has hardly changed when this happens. The ionic dipole relaxes and flips over a longer duration (event D) during which the current settles to its steady state value. Note that the magnitude of various areal densities are plotted in Fig. \ref{Fig4}. Hence the 'notch' in the $Dipole_{Ion}$ curve denotes the transition from mobile ion accumulation to depletion of mobile ions at the ETL side.\vspace{0.5em}\\
Once the current reaches the steady state value (event E), the space charge in ETL is almost exactly balanced by the ionic charge near ETL side. Given the above insights, it is evident that flat band conditions at ETL happens (i.e., ETL band flipping, event C) much before the ionic dipole flips. This is due to the fact that, after event B, $Dipole_{Hole}$  compensate much of the $Dipole_{Ion}$. In addition, the ETL space charge is much smaller in magnitude than the ionic dipole. Accordingly, small decrease in the ionic dipole (not evident in the scale used to plot Fig. \ref{Fig4}) leads to a reversal of ETL bands.\vspace{0.5em}\\
Having described the complex events that define the large signal switching of perovskite devices, let us gather a few interesting aspects: (i) the electron and hole accumulation in the perovskite layer during the transient phase could be much larger than the steady state values (compare the densities at event B and E in Fig. \ref{Fig4}), (ii) the current through the device might appear to have reached steady state while the entire system might be quite far away approaching steady state. 

\section{Discussions}
\begin{figure}[ht]
  \centering
    \includegraphics[width=0.45\textwidth]{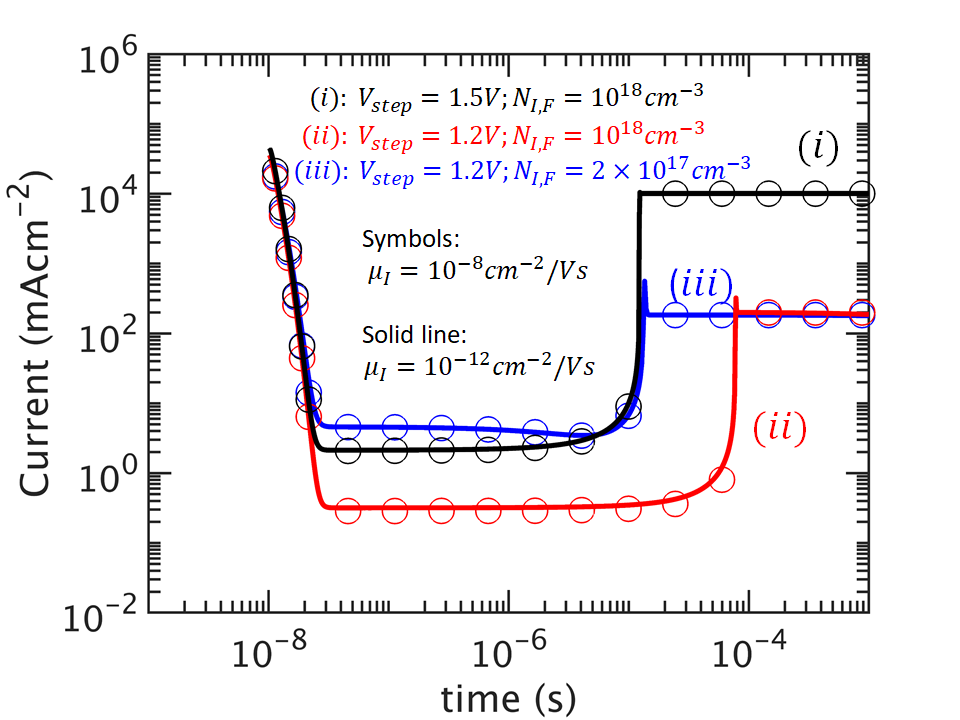}
    \caption{\textit{ Current transient for various combinations of mobile ion density, mobility, and $V_{step}$. Data sets $(i)$ and $(ii)$ have the same ion density but different $V_{step}$ and $\mu_I$, while $(iii)$ differs from $(ii)$ only in terms of ion density. }}
\label{Fig5}
\end{figure}

\textbf{Ion mobility dependence:} Results shown in Fig. \ref{Fig4} indicate that the transition at event B is triggered by the accumulation of electrons and holes. It happens at a timescale where the ion re-distribution is yet to happen. Accordingly, we expect event B to have negligible dependence on ion mobility. Transient simulation results provided in Fig. \ref{Fig5} for two different $\mu_I$, support this inference - i.e., for a given $V_{step}>V_{bi}$ and $N_{I,F}$, current switching is independent of $\mu_I$. Note that the associated time delays are already plotted in Fig. \ref{Fig2}.  As the time delay is independent of ion motion, large signal switching ($V_{step}>V_{bi}$) do not follow the scaling relation given in eq. (1). This resolves the related puzzle on experimental reports on time delays. While it might have been tempting to attribute the mismatch to any apparent differences between drift vs. diffusion mechanism (i.e., through the terms $\Delta V$  and $V_T$ associated with $TD_{diff}$ and $TD_{drift}$, see second paragraph of Introduction), with detailed numerical simulations, here we showed that the switching mechanism is independent of ionic motion.\vspace{0.5em}\\

\textbf{Influence of $V_{step}$:} As before, results shown in Fig. \ref{Fig4} indicates that event B is dictated by the build up of electronic charge in the perovskite layer. Accordingly, larger the $V_{step}$, larger will be the charging current and hence lower will be the TD. Current transients for different $V_{step}$ provided in Fig. \ref{Fig5} supports this inference. Further, the associated TD  plotted in Fig. \ref{Fig2} also indicates the same trends.\vspace{0.5em}\\
\textbf{Influence of ion density:} The ion density has to be large enough to screen the $V_{bi}$ which results in EB profiles shown in Fig. \ref{Fig3}a (else the band diagrams will be that of PIN device with constant electric field under equilibrium conditions). As well known in electrochemistry, the peak density in the ionic dipole is dependent on the bulk ion density. Hence, we expect that event B, which depends on the amount of accumulated charge, will occur at earlier instances with decrease in bulk ion density. The same inference is supported in the results shown in Fig. \ref{Fig5} where current switching happens earlier for $N_{I,F}=2\times 10^{17}cm^2/Vs$ as compared to $N_{I,F}= 10^{18}cm^2/Vs$. We note that current transients for lower $N_{I,F}$  show a decreasing trend after the initial sharp drop (see Fig. \ref{Fig5})  - which is consistent with the literature \cite{doi:10.1021/acsami.9b04991}. \vspace{0.5em}\\
\textbf{Time taken for ionic dipole flipping and steady state:} Our results give fresh insights on the time taken for dipole to flip and achieve steady state. For $\mu_I=10^{-8}cm^2/Vs$, the time taken for ionic dipole to flip and hence steady state conditions is about 0.1s. For $\mu_I=10^{-12}cm^2/Vs$, the corresponding time is of the order of $10^3s$. Hence, accurate estimates for the relevant times indeed scale as per eq. (1). However, it is interesting to note that the time taken by the current to achieve near steady state value is many orders of magnitude lower. Hence, estimates for $mu_I$ based on initial parts of current transients could often be not very reliable. \vspace{0.5em}\\
\textbf{Comparison with literature:}
A direct comparison with experimental results is already provided in Fig. \ref{Fig2}. Our model consistently explains why large signal switching delays do not follow eq. (1). In addition, the switching transients in Fig. \ref{Fig5} compares well with experimental as well as simulation results provided by Neukom et al. \cite{doi:10.1021/acsami.9b04991}. Broad features of our results are consistent with the work by O'Kane et al. \cite{C6TC04964H}. 
Although simulation results were provided in each of the previous two references, we remark that our work stands out in the following aspects - (i) identification $\mu_I$ independent large signal switching across different reports, (ii) explanation of the physical mechanism of large signal switching, especially the collapse of ETL depletion layer and associated charge injection, (iii) provides a clean description of the entire events till steady state, (iv) provides consistent estimates for the time taken to achieve steady state, (v) explains why currents might look invariant although the device has not yet reached steady state conditions, and finally, (vi) all trends were checked against more computationally complex and time consuming simulations using Fermi-Dirac statistics instead of Boltzmann distribution.\vspace{0.5em}\\
\section{Conclusions}
In summary, here we showed, for the first time, that large signal switching of perovskite diode could result in ion mobility independent characteristics. This surprising result is entirely electronic in its origin, while the ionic re-distribution is yet to take place. Our results provides a coherent and consistent explanation for the entire phenomena, and identifies several novel insights in the process. The detailed information provided in this manuscript could help the community to appreciate the ion dynamics in an insightful manner which could lead to better design and optimization of electronic devices based on perovskites.\vspace{-0.7em}\\


\section*{\textbf{Author Information}}
\subsection*{\textbf{Corresponding Authors}}
\noindent Pradeep R. Nair, Department of Electrical Engineering, Indian Institute of Technology Bombay, Mumbai, India. Email: prnair@ee.iitb.ac.in\vspace{-0.7em}\\


\section*{\textbf{Acknowledgements}}
\noindent This project is funded by SERB, DST India. Authors acknowledge IITBNF and NCPRE for computational facilities. PRN acknowledges Visvesvaraya Young Faculty Fellowship.\vspace{-0.7em}\\

\section*{\textbf{Supporting Information}}

\begin{table}[h]
    
    \centering
    
\begin{tabular}{| m{4cm} | m{2cm} | m{1cm} | m{1cm} |}
    \hline
    Parameter, Symbol (Units) & Perovskite & HTL & ETL \\
    \hline
    Thickness of layers, $L_P$, $L_E$, and $L_H$  (nm) & $300$ & $100$ & $100$ \\
    \hline
    Electron Affinity, $\chi$ (eV) & 4 & 2.8 & 4.2\\
    \hline   
    Band gap, $E_g$ (eV) & 1.55 & 2.55 & 2.55\\
    \hline   
    Doping density, $N_D$, $N_A$ ($cm^{-3}$) &  &  $10^{18}$ & $10^{18}$ \\
    \hline
    Density of states, $N_C = N_V$ ($cm^{-3}$) & $10^{19}$ &  $10^{19}$ & $10^{19}$ \\
    \hline
    Ion density, $N_I$ ($cm^{-3}$) & $10^{17}$ - $10^{18}$ & & \\
    \hline
    Ion mobility, $\mu_I$ ($cm^2/Vs$) & $10^{-8}$ - $10^{-12}$ & & \\
    \hline
    Generation rate in the layer, G ($cm^3/s$) & $4.2 \times 10^{21}$ & & \\
    \hline
    SRH lifetime of carriers, $\tau$ ($ns$) & 100 & & \\
    \hline
    Radiative recombination coefficient, ($cm^{-3}/s$) & $ 10^{-10}$& & \\
    \hline
    Mobility of carriers in the layers, $\mu_n = \mu_p$ ($cm^2/Vs$) & 10 & $10^{-3}$ & $10^{-3}$ \\
    \hline
    Relative permittivity of the materials, $\epsilon_r$ & 24 & 4 & 4 \\
    \hline
     \end{tabular}

    \caption{Parameters used in simulations. }
    \label{tab:my_label}
\end{table}
\textbf{Simulation Methodology:}
 The set of equations provided in Section II are self consistently solved to obtain time dependent characteristics. For this, the Poisson's equation is discretized using finite difference scheme while the currents in continuity equations (RHS of eqs 3-5) are evaluated using Scharfatter-Gummel discretization method. The time integration is performed using backward Euler scheme (and the results are verifed through BDF2 method as well). Newton's method is used to iteratively arrive at self-consistent solutions. The results presented were obtained under the assumption of Boltzmann distribution for electrons/holes with abrupt interface between heterojunctions and continuity of quasi-Fermi levels at such interfaces. However, the results were verified against more complex simulations with Fermi-Dirac Statistics through generalized Einstein relation\cite{PierretADF,BLAKEMORE19821067,Nilsson1978}.\\
 The ions are confined to the perovskite layer with appropriate Neumann boundary conditions. Ohmic boundary conditions are assumed at metal contact to P and N regions. 
 \begin{figure}[ht]
  \centering
    \includegraphics[width=0.45\textwidth]{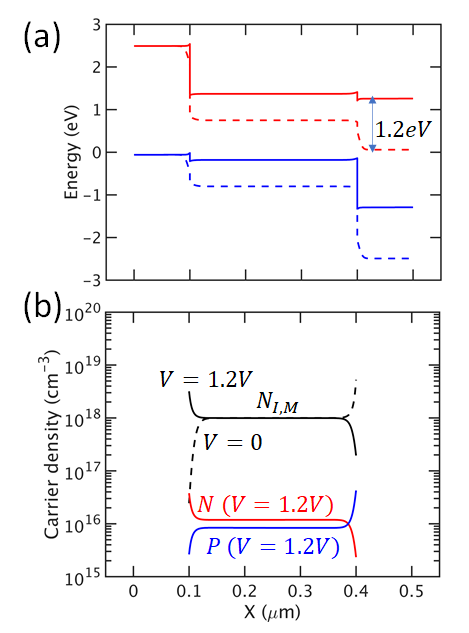}
    \caption{\textit{ Comparison of EB diagram and $N_{I,M}$, $N$ and $P$ profiles in perovskite under equilibrium (dashed lines) and steady state conditions (solid lines) at an applied bias of 1.2V. N and P at equilibrium conditions are of the order of $n_i=10^6 cm^{-3}$ and hence not shown.}}
\label{Fig3}
\end{figure}
 We considered both trap assisted SRH and radiative band-band recombination processes (for the terms $r_N$ and $r_P$ in eqs 3-4 along with $g_N=g_P=0$ for dark conditions). In addition, recombination at ETL/perovskite and HTL/perovskite interfaces is also taken into account through appropriate surface recombination velocities. The initial conditions for time dependent simulations are obtained through steady state solutions for the same set of equations under short circuit conditions (i.e., in dark). Various parameters used are provided in Table 1 of Supplementary materials/Appendix.\vspace{0.5em} \\

We emphasize that our solution methodology follows well established numerical modeling techniques used for heterojunction devices. In addition, the simulation results were calibrated through comparison with commercial packages like Sentaurus and analytical estimates, as appropriate. Table 1 provides a list of various parameters used in simulations.
\textbf{Steady state band diagram}
The band diagrams and carrier density profiles at equilibrium and at $V_{step}=1.2V$ are provided in Fig. \ref{Fig3}.

\begin{figure}[ht]
  \centering
    \includegraphics[width=0.45\textwidth]{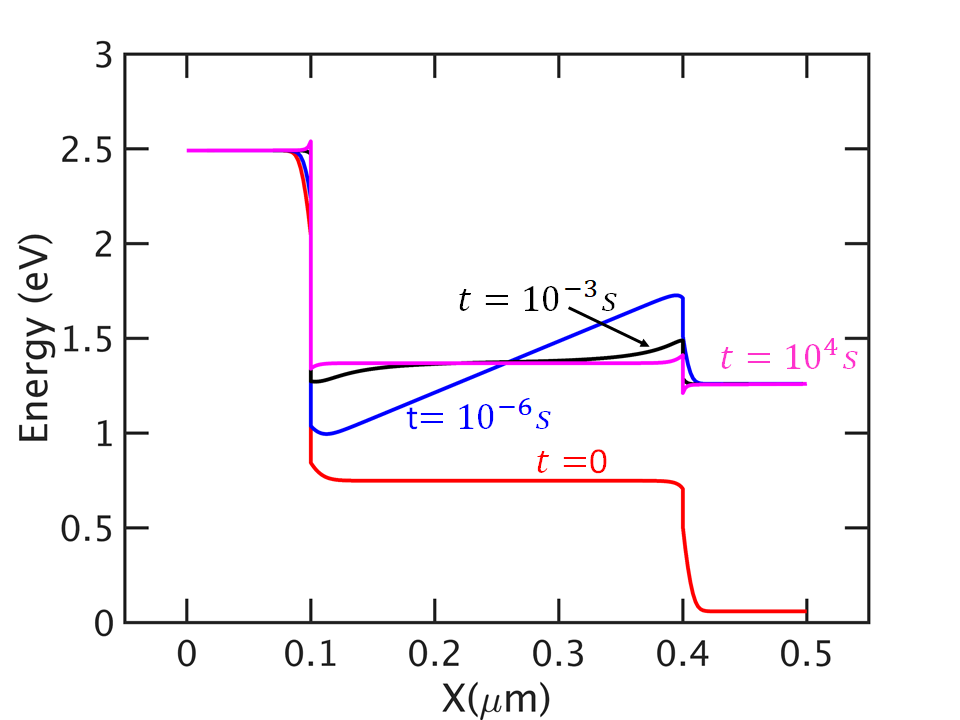}
    \caption{\textit{ Conduction band profile at various time instants corresponding to the results shared in Fig. 4. }}
    \label{FigS2}
    \label{Ec_var}
\end{figure}

\textbf{Time evolution of conduction band}
Fig. \ref{FigS2} shows the variation of conduction band at various critical time instances in the switching transients showed in Fig. \ref{Fig4}. At $t=0$, the no electric field exists in perovskite. At very short time after application of $V_{step}$, the band bending is such that electrons accumulate near HTL while holes accumulate near HTL. After event B (see Fig. \ref{Fig4}), electric field is negligible over most of the perovskite and at long times (event E), the bands reach steady state conditions (which are the same as plotted in Fig. \ref{Fig3}b).\\

\nocite{*}

\bibliography{aipsamp}

\end{document}